\documentclass[]{raa}
\usepackage{graphicx}
\usepackage{times}
\usepackage{natbib}
\usepackage{mathrsfs}
\usepackage{amsmath}
\usepackage{amssymb}
\usepackage{booktabs}
\usepackage{hyperref}

\begin{document}

  \title{On static and spherically symmetric solutions of Starobinsky model}
   \volnopage{Vol. 000 No.0, 000--000}      
   \setcounter{page}{1}          

     \author{Shuang Yu
      \inst{1,2}
   \and Changjun Gao
      \inst{1,2}
   \and Mingjun Liu
      \inst{3}
   }

   \institute{Key Laboratory of Computational Astrophysics, National Astronomical Observatories, Chinese Academy of Sciences, Beijing 100101, China {\it yushuang@nao.cas.cn}\\
        \and
            School of Astronomy and Space Sciences, University of Chinese Academy of Sciences,
No. 19A, Yuquan Road, Beijing 100049, China\\
        \and
          Department of Astronomy, School of Physics, Peking University, Beijing 100871, China\\
          }

\date{Received~~2017 October 16; accepted~~2017~~Feb 22}

\abstract{We investigate the problem of static and spherically symmetric solutions in the Starobinsky gravity model. By extending the Lichnerowicz and Israel theorems, William Nelson have demonstrated that the Schwarzschild solution is the unique static, spherically symmetric, and asymptotically flat black hole solution in the Starobinsky model. However, Hong L{\"u} et al find that there are sign errors in the proof of Nelson. This raises the problem whether Nelson's proof is correct or not. In order to answer this question, we explore the corresponding solutions by using the Taylor series expansion method. We find that Nelson's conclusion is indeed correct despite the flaw in the proof.}


\authorrunning{S. Yu et al.}
\titlerunning{On exact solution of Starobinsky }
\maketitle

\section{Introduction}
\label{sec:1}
The most recent observations from the Planck satellite \citep{Planck:2016} are remarkably consistent with the Starobinsky model of gravity $f(R)=R+sR^2$ \citep{Starobinsky:1980}.
On the contrary, the chaotic inflationary models like large field inflation and natural inflation are disfavored because of their high tensor-to-scalar ratio.
On the other hand, compared to the Starobinsky model, other alternatives of $f(R)$ modified gravity theories are usually plagued by various problems.
For example, they generally lead to the Einstein gravity plus a cosmological constant \citep{Thongkool:2009} or have a field-matter dominated epoch rather than a standard matter-dominated epoch \citep{Amendola:2007} or run into a curvature singularity \citep{Frolov:2008} (see \cite{Kobayashi:2008} for the existence of a singularity in an
asymptotic de Sitter Universe and \citep{Upadhye:2009} for the opposite
statement), or produce high frequency oscillations and a
singularity at finite time in cosmology \citep{Appleby:2008} or give rise to a
fine tuning of the exponent \citep{Faraoni:2011}.
In view of these points, great research enthusiasm on Starobinsky gravity has been caused.

We shall deal with the solution problem of Starobinsky gravity. Actually, the investigations on the static, spherically symmetric spacetime solutions in the other  modified gravity theories have also been carried out.
In the first place, numerical solutions modeled as neutron stars in modified gravities are studied by the following references \citep{Cooney:2010,Babichev:2010,Arapoglu:2011,Jaime:2011,Santos:2012,Orellana:2013,Alavirad:2013,
Astashenok:2013,Ganguly:2014,Astashenok:2014,Yazadjiev:2014,Capozziello:2016,Astashenok:2017}.
Secondly, many efforts are devoted to the aspects of black hole solutions.
Here we are unable to exhaust all the references and we only present a brief review.
Using the Taylor expansion method, the near-horizon geometry of black hole spacetime is investigated in Ref.~\citep{lu:2015}.
The Lifshitz black hole in four dimensional R-squared gravity is presented in Ref.~\citep{Cai:09}.
Some exact, asymptotically non-flat and static black hole solutions are given in Ref.~\citep{Gao:2016}. Static spherically symmetric solutions of $f(R)$ gravity have also been studied in both the vacuum case \citep{Multamaki:2006} and the perfect fluid model \citep{Multamaki:2007}.
Following a perturbative approach in $f(R)$ gravity around the Einstein-Hilbert action,  Ref.~\citep{cruz:2009} found that only solutions of the Schwarzschild-(Anti) de Sitter type are present up to second order in perturbations.
A class of exact static spherically symmetric solution has been presented in \citep{john:05} for the form of $R^{1+\delta}$.
Using the method of Lagrange multiplier, Ref.~\citep{seb:11} presents a Lagrangian derivation of the equation of motion for the static spherically symmetric spacetimes in $f(R)$ theory of gravity.
It is found the corresponding equations of motion are simply of first order derivative and thus some new solutions are obtained.
Ref.~\citep{seb:12} constructs some new static spherically symmetric interior solutions in $R^{1+\delta}$ theory.
With the method of Noether symmetries,  Ref.~\citep{seb:13} finds some new static spherically symmetric solutions in $f(R)$ theory.
Ref.~\citep{Multamaki:2006} constructs the spherically symmetric solutions of $f(R)$ gravity using their input function method.
Ref.~\citep{seb:15} presents a static axially symmetric vacuum solution for $f(R)$ gravity in the Weyl's canonical coordinates.
Finally, using the so-called generator method, Ref.~\citep{ami:15} presents some static spherically symmetric solutions in $f(R)$ theory in $n$ dimensional spacetimes.
For more works on static solutions in $f(R)$ gravity, we refer to the Refs.~\citep{{seb:16,seb:17,seb:18,cruz:14,cruz:16,seb:20,seb:21,seb:22,seb:23,seb:24,seb:25,ser:13,ser:14,hendi:14,Hendi:140,babi:091,hendi:12,rat:15,lu:15prd,chak:2015,fro:2009}}.

In this paper, we shall seek for the static, spherically symmetric and asymptotically flat black hole solution in the Starobinsky model of gravity. In fact, Nelson \citep{nelson:2010} has shown that, by extending the Lichnerowicz and Israel theorems from Einstein gravity to higher derivative gravities, the Schwarzschild solution is the unique static, spherically symmetric,and asymptotically flat black hole solution in the Starobinsky model of gravity. However, Hong L{\"u} et al ~\citep{lu:15prd} find that there are sign errors in the proof of Nelson. This raises the problem whether Nelson's proof is correct or not. In order to answer this question, we explore the corresponding solutions by using the Taylor series expansion method. We find that Nelson's conclusion is indeed correct despite the flaw in the proof.
Compared to Nelson's verification, we can go even further.
Using the Taylor series expansion method, we find the conclusion is applicable not only to the Starobinsky model but also to the more generalized modified gravity, $f(R)=R+s_2R^2+s_3R^3+s_4R^4+O\left(R^4\right)$ (In contrast, Nelson's verification is applicable to the theory of $f(R)=R+sR^2+\alpha R_{\mu\nu}R^{\mu\nu}$). Thus it is an important conclusion.
Throughout this paper, we adopt the system of units in which $G=c=\hbar=1$ and the metric signature $(-,\ +,\ +,\ +)$.
\section{solution in the Einstein gravity}
In this section, as a warm-up, we shall derive the static, spherically symmetric and asymptotically flat spacetime solution in the Einstein gravity by using the Taylor expansion method.

Let's start from the Einstein-Hilbert action
\begin{equation}
S=\int d^4x\sqrt{-g}\frac{R}{16\pi}\;,
\end{equation}
where $R$ is the Ricci scalar. Since we shall look for the vacuum solutions, the contribution of matters is absent in the action.

The metric for static and spherically symmetric spacetime in the isotropic coordinate system can be written as
\begin{equation}
ds^2=-A\left(r\right)^2 dt^2+B\left(r\right)^2\left(dr^2+r^2d\Omega^{2}\right)\;,
\end{equation}
where $d\Omega^2= d\theta^2+\sin^2\theta d\phi^2$.
Then we obtain the Einstein equations
\begin{eqnarray}\label{Einstein}
2BB{''}r-B{'}^2r+4BB{'}=0\;,\label{Einstein1}\\
2BB{'}A{'}r+B{'}^2Ar+2B^2A{'}+2BB{'}A=0\;,\label{Einstein2}\\
A{''}B^2r+BAB{''}r-B{'}^2Ar+B^2A{'}+BB{'}A=0\label{Einstein3}\;.
\end{eqnarray}
The system of equations can be solved and the solution is
\begin{equation}
A=\frac{1-\frac{M}{2r}}{1+\frac{M}{2r}}\;,\ \ \ \ B=\left(1+\frac{M}{2r}\right)^2\;,\label{sch}
\end{equation}
which is the Schwarzschild solution. $M$ is the mass of the source. Different from the Schwarzschild coordinate, the metric components in isotropic coordinate are convergent on the event horizon. So we shall carry out the calculation in the isotropic coordinate ordinate. We expand $A(r)$ and $B(r)$ in the form of Taylor series
\begin{eqnarray}
A(r)&=&1+\frac{a_1M}{r}+\frac{a_2M^2}{r^2}+\frac{a_3M^3}{r^3}+\frac{a_4M^4}{r^4}+\frac{a_5M^5}{r^5}+\frac{a_6M^6}{r^6}+O\left(\frac{1}{r^6}\right)\;,\label{taylor1}\\
B(r)&=&1+\frac{b_1M}{r}+\frac{b_2M^2}{r^2}+\frac{b_3M^3}{r^3}+\frac{b_4M^4}{r^4}+\frac{b_5M^5}{r^5}+\frac{b_6M^6}{r^6}+O\left(\frac{1}{r^6}\right)\;,\label{taylor2}
\end{eqnarray}
where $a_i$ and $b_i$ are dimensionless constants to be determined. For sufficiently large $r$, we have $A=1$ and $B=1$. In other words, the spacetime is asymptotically flat. Substituting Eq.~(\ref{taylor1}) and (\ref{taylor2}) into the Einstein Eqs.~(\ref{Einstein1}), (\ref{Einstein2}) and (\ref{Einstein3}), we find Eqs.~(\ref{Einstein1}), (\ref{Einstein2}) and (\ref{Einstein3}) become
\begin{eqnarray}
0&=&\frac{\alpha_2}{r^2}+\frac{\alpha_3}{r^3}+\frac{\alpha_4}{r^4}+\frac{\alpha_5}{r^5}+\frac{\alpha_6}{r^6}+O\left(\frac{1}{r^6}\right),\label{series1}\\
0&=&\frac{\beta_1}{r}+\frac{\beta_2}{r^2}+\frac{\beta_3}{r^3}+\frac{\beta_4}{r^4}+\frac{\beta_5}{r^5}+\frac{\beta_6}{r^6}+O\left(\frac{1}{r^6}\right),\label{series2}\\
0&=&\frac{\gamma_1}{r}+\frac{\gamma_2}{r^2}+\frac{\gamma_3}{r^3}+\frac{\gamma_4}{r^4}+\frac{\gamma_5}{r^5}+\frac{\gamma_6}{r^6}+O\left(\frac{1}{r^6}\right),\label{series3}
\end{eqnarray}
where $\alpha_i$, $\beta_i$ and $\gamma_i$ are constants whose concrete expressions are determined by $a_i, b_i$ and $M$. We need to find the expressions for twelve coefficients, $a_i$ and $b_i$ in the Eq.~(\ref{taylor1}) and (\ref{taylor2}). So we want twelve algebraic equations.

By setting the leading terms in Eqs.~(\ref{series1}), (\ref{series2}) and (\ref{series3}) to zero, we obtain twelve algebraic equations
\begin{eqnarray}
\alpha_2=0\;, \ \ \ \ \alpha_3=0\;, \ \ \ \ \alpha_4=0\;, \ \ \ \ \alpha_5=0\;, \ \ \ \ \\
\beta_1=0\;, \ \ \ \ \beta_2=0\;, \ \ \ \ \beta_3=0\;, \ \ \ \ \beta_4=0\;, \ \ \ \ \\
\beta_5=0\;, \ \ \ \ \gamma_3=0\;, \ \ \ \ \gamma_4=0\;, \ \ \ \ \gamma_5=0\;. \ \ \ \
\end{eqnarray}
We find $\gamma_1=\beta_1$, $\gamma_2=\beta_2$ and $\gamma_1=0$, $\gamma_2=0$. Solving the twelve equations, we arrive at
\begin{eqnarray}
a_1&=&-b_1\;,\ \ \ \ \ \ \ \ \ \ \ \ \ b_1=b_1\;,\\
a_2&=&\frac{b_1^2}{2}\;,\ \ \ \ \ \ \ \ \ \ \ \ \ \ \ b_2=\frac{b_1^2}{4}\;,\\
a_3&=&-\frac{b_1^3}{4}\;,\ \ \ \ \ \ \ \ \ \ \ \ \ b_3=0\;,\\
a_4&=&\frac{b_1^4}{8}\;,\ \ \ \ \ \ \ \ \ \ \ \ \ \ \ b_4=0\;,\\
a_5&=&-\frac{b_1^5}{16}\;,\ \ \ \ \ \ \ \ \ \ \ \ \  b_5=0\;,\\
a_6&=&\frac{b_1^6}{32}\;,\ \ \ \ \ \ \ \ \ \ \ \ \ \ b_6=0\;.
\end{eqnarray}
We remember $b_1=1$ in the Newtonian limit of Einstein equations. Therefore, we finally have
\begin{eqnarray}
a_1&=&-1\;,\ \ \ \ \ \ \ \ \ \ \ \ \ b_1=1\;,\\
a_2&=&\frac{1}{2}\;,\ \ \ \ \ \ \ \ \ \ \ \ \ \ \ b_2=\frac{1}{4}\;,\\
a_3&=&-\frac{1}{4}\;,\ \ \ \ \ \ \ \ \ \ \ \ \ b_3=0\;,\\
a_4&=&\frac{1}{8}\;,\ \ \ \ \ \ \ \ \ \ \ \ \ \ \ b_4=0\;,\\
a_5&=&-\frac{1}{16}\;,\ \ \ \ \ \ \ \ \ \ \ \ \  b_5=0\;,\\
a_6&=&\frac{1}{32}\;,\ \ \ \ \ \ \ \ \ \ \ \ \ \ b_6=0\;.
\end{eqnarray}
On the other hand, if we expand the Schwarzschild solution, Eq.~(\ref{sch}) using the Taylor series, we have
\begin{eqnarray}
A&=&1-\frac{M}{r}+\frac{M^2}{2r^2}-\frac{M^3}{4r^3}+\frac{M^4}{8r^4}-\frac{M^5}{16r^5}+\frac{M^6}{32r^6}+O\left(\frac{1}{r^6}\right)\;,\\
B&=&1+\frac{M}{r}+\frac{M^2}{4r^2}\;.
\end{eqnarray}
Eqs.~(21-26) are exactly consistent with Eqs.~(27-28). Thus we conclude that there are no problems that we seek for the \emph{exact} solution to any orders by using the Taylor series method.

\section{solution in the Starobinsky model}
In this section, we solve the field equations in Starobinsky model by using the Taylor series method. The action of Starobinsky model in the absence of matters is given by
\begin{equation}
S=\int d^4x \sqrt{-g}\frac{1}{16\pi}\left(R+sR^2\right)\;.
\end{equation}
Here $s$ is a coupling constant which has the dimension of the square of length. The corresponding equations of motion take the form of
 \begin{equation}
G_{\mu\nu}+2sR\left(G_{\mu\nu}+\frac{1}{4}Rg_{\mu\nu}\right)+2s\left(g_{\mu\nu}\nabla^2R-\nabla_{\mu}\nabla_{\nu}R\right)=0\;.
\end{equation}
Given the metric Eq.~(2), the equations of motion are then obtained.
Then we are facing a daunting task of resolving the equations. Therefore, we resort to the Taylor series expansion method.
Substituting  Eqs.~(\ref{taylor1}) and (\ref{taylor2}) into the equations of motion, we obtain
\begin{eqnarray}
0&=&\frac{\alpha_2}{r^2}+\frac{\alpha_3}{r^3}+\frac{\alpha_4}{r^4}+\frac{\alpha_5}{r^5}+\frac{\alpha_6}{r^6}+O\left(\frac{1}{r^6}\right),\label{series11}\\
0&=&\frac{\beta_1}{r}+\frac{\beta_2}{r^2}+\frac{\beta_3}{r^3}+\frac{\beta_4}{r^4}+\frac{\beta_5}{r^5}+\frac{\beta_6}{r^6}+O\left(\frac{1}{r^6}\right),\label{series21}\\
0&=&\frac{\gamma_1}{r}+\frac{\gamma_2}{r^2}+\frac{\gamma_3}{r^3}+\frac{\gamma_4}{r^4}+\frac{\gamma_5}{r^5}+\frac{\gamma_6}{r^6}+O\left(\frac{1}{r^6}\right),\label{series31}
\end{eqnarray}
where $\alpha_i$, $\beta_i$ and $\gamma_i$ are determined by $a_i\, b_i$ and $M$. In this case, we also have $\gamma_1=\beta_1$ and  $\gamma_2=\beta_2$. Thus we have the following algebraic equations
 \begin{eqnarray}
\alpha_2=0\;, \ \ \ \ \alpha_3=0\;, \ \ \ \ \alpha_4=0\;, \ \ \ \ \alpha_5=0\;, \ \ \ \ \\
\beta_1=0\;, \ \ \ \ \beta_2=0\;, \ \ \ \ \beta_3=0\;, \ \ \ \ \beta_4=0\;, \ \ \ \ \\
\beta_5=0\;, \ \ \ \ \gamma_3=0\;, \ \ \ \ \gamma_4=0\;, \ \ \ \ \gamma_5=0\;. \ \ \ \
\end{eqnarray}
Solving these equations, we obtain
\begin{eqnarray}
a_1&=&-b_1\;,\ \ \ \ \ \ \ \ \ \ \ \ \ b_1=b_1\;,\\
a_2&=&\frac{b_1^2}{2}\;,\ \ \ \ \ \ \ \ \ \ \ \ \ \ \ b_2=\frac{b_1^2}{4}\;,\\
a_3&=&-\frac{b_1^3}{4}\;,\ \ \ \ \ \ \ \ \ \ \ \ \ b_3=0\;,\\
a_4&=&\frac{b_1^4}{8}\;,\ \ \ \ \ \ \ \ \ \ \ \ \ \ \ b_4=0\;,\\
a_5&=&-\frac{b_1^5}{16}\;,\ \ \ \ \ \ \ \ \ \ \ \ \  b_5=0\;,\\
a_6&=&\frac{b_1^6}{32}\;,\ \ \ \ \ \ \ \ \ \ \ \ \ \ b_6=0\;.
\end{eqnarray}
They are exactly for the Schwarzschild solution when $b_1=1$. We note that the coupling constant $s$ does not emerge in the expressions of the coefficients. This means the Ricci-squared term $sR^2$ makes no contribution in the equations of motion. This observation is consistent with the proof of Nelson. In fact, Nelson showed that any
static black-hole solution of Starobinsky model must have vanishing Ricci scalar. Taken this into consideration, the equations of motion Eq.~(30) is simply
\begin{eqnarray}
G_{\mu\nu}=0\;.
\end{eqnarray}
Thus the Ricci-squared term $sR^2$ does make no contribution to the equations of motion. Up to this point, we confirm that Nelson's conclusion \emph{the unique, static, spherically symmetric and asymptotically flat black hole solution in the Starobinsky model is nothing but the Schwarzschild solution} is correct, despite the flaw in his proof. Compared to Nelson's verification, we can go further. Using the Taylor series expansion method, we have checked that for more general modified gravity
\begin{equation}
S=\int d^4x \sqrt{-g}\frac{1}{16\pi}\left[R+s_2R^2+s_3R^3+s_4R^4+O\left(R^4\right)\right]\;,
\end{equation}
the unique, static, spherically symmetric and asymptotically flat black hole solution remains the Schwarzschild solution. Then the applicability of the conclusion is extended. It should be noted that Canate recently presents a no-hair theorem which discards the existence of four dimensional asymptotically flat, static and spherically symmetric or stationary axial symmetric, non-trivial black holes in the frame of $f(R)$  gravity \cite{ped:2016}. This is a huge support for our examination.

\section{conclusion and discussion}
In conclusion, we have investigated the solution problem in the Starobinsky gravity model. By using the Taylor series expansion method, we show the static, spherically symmetric and asymptotically flat spacetime is uniquely the Schwarzschild solution. This is in agreement with the proof of Nelson who uses the method of extending the Lichnerowicz and Israel theorems from Einstein gravity to higher order theories of gravity, despite the flaw in his proof. But compared to Nelson's verification, we can go further. We find the conclusion can be extended the more generalized modified gravity. Thus it is an important conclusion.

We can also understand the conclusion in another way of thinking. We remember that the $f(R)$ theory of modified gravity is equivalent to the scalar tensor theory
\begin{equation}
\tilde{S}=\int d^4x \sqrt{-\tilde{g}}\left[\frac{\tilde{R}}{16\pi}+\frac{1}{2}\nabla_\mu\tilde{\phi}\nabla^{\mu}\tilde{\phi}+V\left(\tilde{\phi}\right)\right]\;,
\end{equation}
by making variable redefinitions and conformal transformations \cite{sot:10}:
\begin{equation}
g_{\mu\nu}=e^{-\sqrt{\frac{16\pi}{3}}\tilde{\phi}}\tilde{g}_{\mu\nu}\;.
\end{equation}
Refs.~\cite{haw:72,Sudarsky:1995} have shown that black holes in the above scalar tensor theory cannot carry scalar charge for arbitrary potentials. This is known as black-hole no-hair theorem.  The theorem guarantees the scalar field $\tilde{\phi}$ is constant in this spacetime. Then the scalar potential plays the role of the cosmological constant. The no-hair theorem requires the Schwarzschild-de Sitter solution is the unique static spherically symmetric spacetime. If one assume the spacetime is asymptotically flat, we are left with only the Schwarzschild solution $\tilde{g}_{\mu\nu}$. From Eq.~(46) we see the solution $g_{\mu\nu}$ of $f(R)$ gravity is exactly the Schwarzschild solution since it is proportional to $\tilde{g}_{\mu\nu}$ by a constant. In all, we have good reason to believe that \emph{the Schwarzschild solution is the unique static, spherically symmetric and asymptotically flat spacetime solution in the Starobinsky gravity.}

\section*{Acknowledgments}
This work is partially supported by the Strategic Priority Research
Program ``Multi-wavelength Gravitational Wave Universe'' of the
CAS, Grant No. XDB09000000 and XDB23040100, and the NSFC
under grants 10973014, 11373020, 11465012 and 11633004 and
the Project of CAS, QYZDJ-SSW-SLH017.

\bibliographystyle{mnras}
\bibliography{reference}

\end{document}